# Information flow between resting state networks

Ibai Diez[a,*], Asier Erramuzpe[a,*], Iñaki Escudero[a,b], Beatriz Mateos[a,b], Alberto Cabrera[c], Daniele Marinazzo[d], Ernesto J. Sanz-Arigita[e], Sebastiano Stramaglia[f,g] and Jesus M. Cortes[a,h,i**]

and for the Alzheimer's Disease Neuroimaging Initiative[1]

a. Biocruces Health Research Institute. Cruces University Hospital, Plaza de Cruces S/N, 48903, Barakaldo, Bizkaia, Spain
b. Radiology Service. Cruces University Hospital, Plaza de Cruces S/N, 48903, Barakaldo, Bizkaia, Spain
c. Osatek. C/ Alava, nº45, 01006 Vitoria – Gazteiz, Alava, Spain
d. Faculty of Psychology and Educational Sciences, Department of Data Analysis, Ghent University, Henri Dunantlaan 1 B-9000 Gent , Belgium
e. Radiology and Image Analysis Center, VUmc, De Boelelaan 1117, 1081 HZ Amsterdam, The Netherlands
f. Dipartimento di Fisica, Universita degli Studi di Bari and INFN, Via G. Amendola 173, 70126 Bari, Italy
g. BCAM - Basque Center for Applied Mathematics, Alameda Mazarredo 14, 48009, Bilbao, Spain
h. Ikerbasque, The Basque Foundation for Science. Alameda Urquijo, 36-5 Plaza Bizkaia, 48011 Bilbao, Bizkaia, Spain
i. Departamento de Biologia Celular e Histologia. University of the Basque Country. Barrio Sarriena s/n, 48940 Leioa, Bizkaia, Spain

[1]Data used in the preparation of this article were obtained from the Alzheimer's Disease Neuroimaging Initiative (ADNI) database (adni.loni.ucla.edu). As such, the investigators within the ADNI contributed to the design and implementation of ADNI and/or provided data but did not participate in the analysis or writing of this report. A complete listing of ADNI investigators is available at http://adni.loni.usc.edu/wp-content/uploads/how_to_apply/ADNI_Acknowledgement_List.pdf

*The two authors have contributed equally

**Correspondence to:
Jesus M Cortes Diaz
Computational Neuroimaging Lab.
Biocruces Health Research Institute.
Cruces University Hospital,
Plaza de Cruces S/N, E-48903,
Barakaldo, Bizkaia, Spain
Tel +34 94 600 6000 (ext. 5199)
E-mail: jesus.cortesdiaz@osakidetza.net

**Abstract**

The resting brain dynamics self-organizes into a finite number of correlated patterns known as resting state networks (RSNs). It is well known that techniques like independent component analysis can separate the brain activity at rest to provide such RSNs, but the specific pattern of interaction between RSNs is not yet fully understood. To this aim, we propose here a novel method to compute the information flow (IF) between different RSNs from resting state magnetic resonance imaging. After hemodynamic response function blind deconvolution of all voxel signals, and under the hypothesis that RSNs define regions of interest, our method first uses principal component analysis to reduce dimensionality in each RSN to next compute IF (estimated here in terms of Transfer Entropy) between the different RSNs by systematically increasing k (the number of principal components used in the calculation). When $k = 1$, this method is equivalent to computing IF using the average of all voxel activities in each RSN. For $k \geq 1$ our method calculates the k-multivariate IF between the different RSNs. We find that the average IF among RSNs is dimension-dependent, increasing from $k = 1$ (i.e., the average voxels activity) up to a maximum occurring at $k = 5$ to finally decay to zero for $k \geq 10$. This suggests that a small number of components (close to 5) is sufficient to describe the IF pattern between RSNs. Our method -- addressing differences in IF between RSNs for any generic data -- can be used for group comparison in health or disease. To illustrate this, we have calculated the inter-RSNs IF in a dataset of Alzheimer's Disease (AD) to find that the most significant differences between AD and controls occurred for $k = 2$, in addition to AD showing increased IF w.r.t. controls. The spatial localization of the $k = 2$ component, within-RSNs, allows the characterization of IF differences between Alzheimer's and controls.

**List of acronyms:**

RSN: Resting State Network

rs-fMRI: Resting State Functional Magnetic Resonance Imaging

IF: Information Flow

ADNI: Alzheimer's Disease Neuroimaging Initiative

DMN: Default Mode Network

AD: Alzheimer's Disease

ROI: Region Of Interest

ICA: Independent Component Analysis

PLS: Partial Least Squares

PCA: Principal Components Analysis

DFC: Directed Functional Connectivity

DCM: Dynamic Causal Modeling

BOLD: Blood Oxygen Level Dependent

GC: Granger Causality

TE: Transfer Entropy

NIA: National Institute on Aging

NBIB: National Institute of Biomedical Imaging

FDA: Food and Drug Administration

TR: Repetition Time

CSF: Cerebrospinal Fluid

TFCE: Threshold-Free Cluster Enhancement

AAL: Automated Anatomical Labeling

HRF: Hemodynamic Response Function

MRI: Magnetic Resonance Imaging

PET: Positron Emission Tomography

**Introduction**

The overall brain dynamics generated at rest can be decomposed as a superposition of multiple activation patterns, the so-called Resting State Networks (RSNs). Becoming a fundamental characteristic of brain function, RSNs are a pivotal element for understanding the dynamics and organization of the brain basal activity in health and disease [Raichle et al., 2001, Fox et al., 2005, Raichle and Mintum, 2006, Raichle and Snyder, 2007, Raichle, 2009]. RSNs emerge from the correlation in signal fluctuations across brain regions during resting state, a condition defined by the absence of goal-directed behavior or salient stimuli. Despite the simplicity of the context in which these brain activity patterns are generated, RSNs dynamics is rich and complex. Different RSNs have been associated to specific cognitive networks e.g., there are visual networks, sensory-motor networks, auditory networks, default mode networks, executive control networks, and some others (for further details see for instance [Beckmann et al., 2005] and references therein).

Currently, it is well-established that a number of techniques, like Region-Of-Interest approach (ROI; seed-based), Independent Component Analysis (ICA) and Partial Least Squares (PLS), can decompose the resting state functional magnetic resonance images (rs-fMRI) to provide such RSNs [Bell and Sejnowski, 1995, Beckmann and Smith, 2003, Beckmann et al., 2005, McIntosh and Lobaugh, 2004, Zhang and Raichle, 2010]. Using these techniques, altered functional connectivity in specific RSNs has been described in brain pathological conditions such as in patients with deficit of consciousness after traumatic brain injury [Boveroux et al., 2010, Noirhomme et al., 2010, Heine et al., 2012, Maki-Marttunen et al., 2013], schizophrenia

[Woodward et al., 2011, Karbasforoushan and Woodward, 2012] and epilepsy [Liao et al., 2010]. In the particular case of Alzheimer's Disease (AD), the pathology we have addressed here, starting from the pioneer contribution showing alterations in fMRI [Li et al., 2002], thereon it has been reported a decrease of functional connectivity in the Default Mode Network (DMN, a RSN related to memory function) and in the salience network at both early and advanced stages of the AD disease [Greicius et al., 2004, Rombouts et al., 2005, Binnewijzend et al., 2012, Sheline and Raichle, 2013].

Despite this emphasis on specific networks, it is important to realize that, however, the RSNs functional-division into separate systems does not imply that the brain activity is composed of functional networks working in isolation [Damoiseaux et al., 2006]. Contrarily, the brain regions underlying the inter-RSNs relations have their independent organization and have a different function to the specific ones that each individual RSN has (for recent reviews see [Fornito et al., 2013] and [Smith et al., 2013]). These interactions between brain regions belonging to different RSNs have been described by means of whole-brain connectivity analysis techniques like graph analysis [Sporns et al., 2004, Ponten et al., 2007, Stam and Reijneveld, 2007, Bullmore and Sporns, 2009], and more recently, where connectivity was also analyzed during different cognitive tasks conditions [McLaren et al, 2014].

However, the analysis and interpretation of information flow (IF) between the RSNs remains an open question and has given rise to two contrasting theories attempting to interpret the resting brain activity. The functional integration theory (for review, see [Zhang and Raichle, 2010]) proposes that brain activity during resting-state requires the coordinated activity of all brain RSNs to support the reconstruction, analysis and simulation of experiences or possible scenarios to provide adaptive behavioral advantage. On the other hand, the functional segregation theory states that modality-specific mental activity (eg., image- or language-based thoughts) is related to functional disconnection between the brain networks active during rest [Delamillieure et. al., 2010]. Therefore, resting activity would be maintained by functionally segregated inner-oriented and sensory-related cognition and the respective intrinsic and extrinsic brain networks they depend on.

The question of how different RSNs speak to each other is particularly relevant as it has been reported the existence of disease-driven changes in functional connectivity between different brain networks [Sanz-Arigita et al., 2010, Schoonheim et al., 2013]. Importantly, as the disease progresses responsible RSNs are in turn affected, and this gradual loss of functional connectivity within networks is accompanied by a loss of functional correlation between them [Brier et al., 2012]. This indicates that altered functional activity induced by a local dysfunction might influence the functioning of additional brain regions, leading to the spread of changes in brain activity beyond the network originally affected and in turn the whole-brain functional connectivity pattern [Sanz-Arigita et al., 2010].

Here we propose a new method to study RSNs inter-communication. The method is based on the inference of IF in terms of Transfer Entropy (TE) to study Directed Functional Connectivity (DFC) between RSNs[1]. We apply this method to AD (with the aim of showing one potential application, but remark that the method is general and can be applied to any other disease). The first step is to consider RSNs to function as spatial templates, i.e. masks, similar to the ones reported in [Beckmann et al., 2005]. Similar approaches, considering the RSNs masks rather the independent components per se, have been used before in previous work [Tagliazucchi et al., 2012; Carhart-Harris et al., 2012; Haimovici et al., 2013; Tagliazucchi et al., 2014]. Next, we extracted all the time-series of rs-fMRI activity belonging to each RSN. After hemodynamic response function (HRF) blind deconvolution of all activity signals, our method does not work out with the average of all voxel activities per RSN, rather it approximates the global RSN activity by a k-components signal obtained by Principal Components Analysis (PCA). The IF between regions is thus evaluated as the number of components k is varied, whilst the complexity of the model is controlled by statistical testing. The standard TE analysis is recovered when (i) no HRF deconvolution is made and (ii) just the first principal component is used to describe each RSN (which is equivalent to make the average over voxels within that RSN).

---

[1] For brain connectivity studies, there exist different approaches for DFC [Granger, 1969, Schreiber, 2000, Friston et al., 2003, Penny et al., 2004, Roebroeck et al., 2005, Barnett et al., 2009, Marinazzo et al., 2011, Friston, 2011, Bressler and Seth, 2011, Friston, 2009]. One possibility for calculating DFC from fMRI time-series is Dynamic Causal Modeling (DCM), addressing how the activity in one brain area is affected by the activity in another area using explicit models of effective connectivity (for details see for instance [Friston et al., 2003, Penny et al., 2004]). Alternatively, data-driven approaches for DFC work directly with the time-series, and not any further assumption have to be taken, neither about the hemodynamic response, nor about the biophysics from individual neuron to Blood Oxygen Level Dependent (BOLD) level. Two popular data-driven methods for calculating DFC are Granger causality (GC) [Granger, 1969] and transfer entropy (TE) [Schreiber, 2000]; for the Gaussian approximation the two methods are equivalent [Barnett et al., 2009].

To show one potential application of this method, we apply it to a dataset of AD patients from the Alzheimer's Disease Neuroimaging Initiative (ADNI) and compared the results of inter-RSNs communication with a group of healthy subjects.

**Material and Methods**

*ADNI*

Data used in the preparation of this article were obtained from the ADNI database (adni.loni.ucla.edu). ADNI was launched in 2003 by the National Institute on Aging (NIA), the National Institute of Biomedical Imaging and Bioengineering (NIBIB), the Food and Drug Administration (FDA), private pharmaceutical companies and non- profit organizations, as a $60 million, 5-year public-private partnership. The primary goal of ADNI has been to test whether serial MRI, PET, other biological markers, and clinical and neuropsychological assessment can be combined to measure the progression of Mild Cognitive Impairment (MCI) and early AD. Determination of sensitive and specific markers of very early AD progression is intended to aid researchers and clinicians to develop new treatments and monitor their effectiveness, as well as lessen the time and cost of clinical trials.

The Principal Investigator of this initiative is Michael W. Weiner, MD, VA Medical Center and University of California - San Francisco. ADNI is the result of efforts of many co-investigators from a broad range of academic institutions and private corporations, and subjects have been recruited from over 50 sites across the U.S. and Canada. The initial goal of ADNI was to recruit 800 subjects but ADNI has been followed by ADNI- GO and ADNI-2. To date, these three protocols have recruited over 1500 adults, ages 55 to 90, to participate in the research, consisting of cognitively normal older individuals, people with early or late MCI, and people with early AD. The follow up duration of each group is specified in the protocols for ADNI-1, ADNI-2 and ADNI-GO. Subjects originally recruited for ADNI-1 and ADNI-GO had the option to be followed in ADNI-2. For up-to-date information, see www.adni-info.org.

*Subjects*

The analysis was performed on n = 10 healthy subjects as control (5 males, 5 females, 73.70 ± 1.16 years old) and n = 10 AD patients (5 males, 5 females, 73.40 ± 1.08 years old) and both data sets were downloaded from the ADNI database.

Notice that rather than increasing the population size to a very large number, we preferred to select two small populations but choosing them the most balanced as possible with regard to age and gender. Demographic data (including the ADNI identifier) are given in Tables 1 and 2.

*MRI acquisition and preprocessing*

High resolution anatomical scans and T2 weighted resting state fMRI data were used from each subject. For the fMRI data a total of 140 volumes were acquired with a repetition time (TR) of 3000ms and 64 x 64 matrix with 48 oblique axial slices (voxel size: 3.3125 x 3.3125 x 3.3125mm). The fMRI preprocessing was performed using FSL (FMRIB Software Library v5.0) and AFNI [Cox, 1996]. Data were motion corrected and smoothed using a Gaussian Kernel of 6 mm FWHM. After intensity normalization, a low-pass filter was applied within the slow fluctuations range of (0.01-0.1 Hz) that characterizes the resting state BOLD activity. Next, linear and quadratic trends were removed. Finally motion time courses, white matter signal, cerebrospinal fluid (CSF) signal and global signal were regressed out from the data.

*ROIs definition from RSNs masks*

We defined each ROI as the voxels belonging to each RSN, by using the masks reported in [Beckmann et al., 2005]. These masks can be downloaded from http://www.fmrib.ox.ac.uk/analysis/royalsoc8/. Notice that we are not dealing with the independent components per se, but with the multivariate activity of all the voxels time-series localized within the masks. Specifically, we have used the following eight RSNs: medial visual, lateral visual, auditory, sensory-motor, default mode, executive control, dorsal visual right, dorsal visual left. fMRI data was transformed to the MNI152 template (at 3mm*3mm*3mm resolution) and all the time series from the voxels belonging to each RSN where extracted to define each ROI. The size for each ROI is given in Table 3.

*HRF blind deconvolution*

We individuated point processes corresponding to signal fluctuations with a given signature and extracted a voxel-specific HRF to be used for deconvolution, after following an alignment procedure. The parameters for blind deconvolution were chosen with a physiological meaning, according to [Wu et al. 2013]: for a TR equal to 3s, the threshold was fixed to 1 SD (standard deviation) and the maximum time lag varying from 3 to 5 TR, but results did not change. Results in the manuscript have been done for maximum time lag equal to 5 TR. For further details on the complete HRF blind deconvolution method and the different parameters to be used, see [Wu et. al., 2013]. The resulting time-series after the HRF blind deconvolution are the ones used for the calculation of the IF.

*Information Flow between RSNs*

Let us consider two RSNs A and B. We use TE to estimate IF from A to B and vice-versa; TE is equivalent to Granger Causality (GC) in the Gaussian case [Barnett et al., 2009], which is the case considered here. Let A be described by $N_A$ continuous time-series $\{x_\alpha^A(t)\}_{\alpha=1,...,N_A}$ (ie. set of voxels belonging to region A), and B described by $N_B$ continuous time series $\{x_\alpha^B(t)\}_{\alpha=1,...,N_B}$. All the time series have been de-convolved by the HRF. Fixing the number k of the principal components the same for A and B, we represent the dynamics for A as $\{y_i^A(t)\}_{i=1,...,k}$ and the one for B as $\{y_i^B(t)\}_{i=1,...,k}$. Note that these components are statistically independent in the Gaussian approximation. Given the order m (the number of past points to be included in the state vector) and the lag parameter $\delta$, we denote for the ith-component the state vector

$$Y_i^A(t) = \left(y_i^A(t-\delta) \cdots y_i^A(t-\delta-m)\right), \qquad \text{(Eq. 1)}$$

and the corresponding one for B

$$Y_i^B(t) = \left(y_i^B(t-\delta) \cdots y_i^B(t-\delta-m)\right). \qquad \text{(Eq. 2)}$$

According to Akaike information criterion, we use m=1. The calculation of IF from A to B

depends on the used lag δ. Here, in addition to lag equal to 1, we also simulated δ=2 (corresponding to 6 seconds as TR=3s) and δ=3 (9 seconds), showing slightly different behavior. In particular, for both δ=2 and δ=3, we still found an increase of IF in AD w.r.t. controls, but that increment was not significant in any number of principal components (k). Thus, only for δ=1, the case we have considered here, the increment of IF occurring in AD was statistically significant (for both k=2 and k=3).

In order to evaluate the IF from A to B, we first calculate the IF from A to the ith-component of B, ie.,

$$t(A \rightarrow B^i) = H(y_i^B | \{Y_j^B\}_{j=1,\ldots,k}) - H(y_i^B | \{Y_j^B\}_{j=1,\ldots,k}; \{Y_j^A\}_{j=1,\ldots,k}), \qquad \text{(Eq. 3)}$$

where H is the conditional Shannon entropy, evaluated over the empirical distribution of samples at hand under the assumption of gaussianity, see [Barnett et al, 2009]. Remark two important issues from eq. (3): first, that the two condionated states $\{Y_j^B\}_{j=1,\ldots,k}$ and $\{Y_j^A\}_{j=1,\ldots,k}$ are accounting for lagged interactions, ie., the two depend on the δ parameter appearing in eqs. (1) (2), but the term $y_i^B$ is not lagged. Second, that the interaction given by eq. (3) is univariate for the target and multivariate for the driver.

Next, repeating the same as in eq. (3) for each of the components in B, ie., i=1,…,k, and denoting $\pi_i$ as the probability of $t(A \rightarrow B^i)$ under the null hypothesis of the absence of influence, both $\pi_i$ and $t(A \rightarrow B^i)$ can be calculated analytically in the Gaussian approximation –cf. eq. (12)

in paper [Barnett et al., 2009]–. Finally, IF is then estimated as the average over all components in B, i.e.,

$$IF(A \rightarrow B) = \frac{1}{k}\sum_{i=1}^{k} \theta\left(\frac{\tau}{k} - \pi_i\right) t(A \rightarrow B^i), \qquad (Eq.\ 4)$$

where $\theta$ is the Heaviside function, what makes the sum to account for all the contributions which are statistically significant according to Bonferroni criterion, i.e. those with $\pi_i < \tau/k$, where $\tau$ is the statistical significance (we use 0.05 here).

The complexity of the model is thus controlled by statistical testing, i.e. accepting only significant interactions (a similar strategy to control complexity is used in [Marinazzo et al., 2008]).

*Statistical testing*

To perform statistical significance between groups in figures 2-4, a non-parametric Wilcoxon rank sum test was used to validate the hypothesis that two data distributions have equal medians. This was implemented in Matlab, The Mathworks, Inc., with the function *ranksum* at p=0.05 with Bonferroni correction.

For figure 5, a non-parametric two-sample unpaired t-test was performed as implemented in FSL. First, the k = 2 component was regressed into 4D fMRI data of each particular subject to obtain a subject-specific spatial map. Next, to search for the group differences (control vs AD) in the spatial maps we performed a permutation-based nonparametric inference as implemented in the function *randomise* in FSL, option Threshold-Free Cluster Enhancement (TFCE) with FWE-Corrected p=0.05.

*Anatomical localization in brain differences: control vs AD*

To localize the spatial maps plotted in figure 5, we overlapped these maps with the Automated Anatomical Labeling (AAL) atlas [Tzourio-Mazoyer et al., 2002] to get the anatomical regions underlying such differences. In particular, we calculated the overlapping percentage between all voxels in each spatial map and each of the 45 homologue brain areas existing in the AAL parcellation. Although the RSNs are widespread across the whole brain, our localization criteria only considered percentage of voxels in each spatial map to be bigger than 15% (with respect to the total spatial map size), results given in Table 4.

**Results**

After extracting all voxels belonging to each RSN and de-convolving each of the individual voxels with HRF (see methods), we first applied principal components analysis to each RSN and then computed the multivariate IF between RSNs (see figure 1 for a chartflow). Regarding the amount of variability captured by the principal components, after averaging between subjects, about 60% of total variability was captured by the 10 first components (k=10) (figure S1) and a 99% of total variability was captured for k=72 (figure S2). No statistically significant differences were found in the amount of variability between control and AD, indicating that the data representation in the different principal components were disease independent. Therefore, the reader should not be confused and be aware that differences in IF are not related to these percentages.

The average IF between the different RSNs is represented in figure 2. Systematically, we recomputed IF with a different number of principal components (maintaining the same for all RSNs) from k = 1 up to k = 15. The case k = 1 is equivalent to calculating IF between the average voxel activities for each RSN. The case k > 1 corresponds to a multivariate situation. For visualization purposes, figure 2 shows results up to k = 11 as for k ≥ 10 the information was zero. Notice that to have a zero IF is possible because, according to eq. (4), the quantity IF is averaged over all the components. Thus, if adding a new component does not provide new independent information, the term (1/k) in the denominator will eventually decrease the average IF. To represent figure 2, notice that as we were dealing with n = 10 healthy subjects and n = 10 AD patients, we first obtained for each subject a matrix of IFs in which the element (i, j) indicated IF from the ith-RSN to the jth-one. Next, we pooled together all the matrices belonging to the same group (control and AD) and represented the average IF among all possible pairs --for 8 RSNs, the total number of pairs is 8 * (8 − 1) = 56, which is equal to the number of pairs minus the elements in the principal diagonal--. The profile of average information as a function of the number of principal components (used for calculation of the IF) is the same for control and AD: starting to increase from k = 1 up to the maximum at k = 5 to monotonically begin to decrease. Therefore we conclude that for dimensions bigger than k = 5, a higher dimension does not provide more IF. Statistically significant differences for the average IF between control and AD occurred for k = 2 and k = 3. Interestingly, the average inter-RSNs IF is higher in AD than in controls.

Figure 3 shows all possible values of IF between all RSNs. Here, the number of principal components was fixed to k = 2 (the one with biggest statistical difference in figure 2), but similar graphs were obtained for each value of k. Unlike correlations, TE measures directed functional connectivity, thus figure 3 is representing for each condition control and AD the two directions in IF (ie., for two generic RSNs A and B, if in top panel we represented IF from A to B, then in bottom panel we depicted IF from B to A). In other words, the average value among all the flows in figure 3 is the one plotted in figure 2. It is important to remark that all RSNs are communicating each to another, and, as it is shown figure 2, in average, the AD condition had higher information flowing between RSNs in comparison to control.

Next, we addressed IF arriving to and originating from each specific RSN. Figure 4 shows the outward information in blue and the inward information in red. The negativity in all the bars (obtained by subtracting the information in control minus the corresponding one in AD) confirmed that within all RSNs there existed an increase of the information for AD in both the outward and the inward directions.

Finally, we performed a two-sample unpaired t-test to localize the differences between AD and control for the k = 2 component. Results are showed in figure 5; the differences are plotted in two colors: blue for the activity existing in control but nonexistent in Alzheimer's and in red, vice versa, activated areas belonging to Alzheimer's but not to control. Next, we overlapped these maps with the AAL atlas (details in Methods) to get the relevant anatomical regions for each RSN, cf. Table 4.

**Discussion**

RSNs are chiefly characterized by their universal emergence, meaning that, beyond individual subject differences, RSNs are ubiquitous in healthy brains. Whilst the emergence of RSNs in health is a well-known fact, how these RSNs speak to each other is not fully understood yet. In our approach, instead of applying ICA to each subject separately to get their specific RSNs, we used templates provided in [Beckmann et al., 2005] for all subjects. The use of the same templates for all the subjects is indeed assuming that the spatial structure of all RSNs is universal. Our main hypothesis here is that the breakdown of this assumption might differentiate healthy subjects and AD patients --but the same method can be applied to any other disease-- and that searching for such differences might provide further insights about the alteration patterns of IF in the pathological brain. In other words, we were interested in providing an answer to the following question: what are the different features in healthy vs pathological brains under the hypothesis that the same templates characterize RSNs? To this end, we introduced a novel method that calculates IF between all the different RSNs.

We applied this methodology to AD datasets and compared these results to control. The interesting answer is that such assumption led --for the particular situation of AD-- to differences in the information related to the second and third principal components and not to the first one (which is coincident with the average activity over all voxels per RSN). Notice that from eqs. (3) and (4) one can see that the TE is calculated univariate for the target and multivariate for the driver. Thus, when we say that AD vs. control differences are associated to $k=2$, we mean a multivariate driver having into account the two time series $k=1$ and $k=2$.

Therefore, as long as the average time series of each ROI is concerned, no differences between the patterns of healthy and AD patients emerged. Performing a paired t-test to spatially localize the k = 2 component, we found the regions which were underlying those differences.

An important limitation of ROI analysis in MRI data is the huge number of voxels determining each brain region. In most cases, the average time series (across voxels at each time-point) or the first principal component is assumed to be the ROI representative (we verified that the average and the first principal components are equivalent on our data set); the connectivity analyses are then carried out using these regional representative signals. However, neither the average nor the 1st PCA are taking into account the predictive power of past value. Thus, relevant temporal information may be diluted when considering these signals.

Some approaches have represented each ROI using more than one time series, either using PCA [Zhou et al., 2009] or cluster analysis [Sato et al., 2010]. The approach we are presenting here is rooted on principal component analysis and its novelty is based on two points: first, we preprocessed the time series corresponding to each individual voxel by HRF blind deconvolution and second, instead of fixing the number of components according to a prescribed fraction of the data variance, as it is usually done, we have analyzed IF as the number of components increases, including more details of the ROIs dynamics.

Using the RSNs spatial templates (masks) reported in [Beckmann et al., 2005], we extracted different ROIs to approximate each RSN by a fixed multivariate dimension (k) found by PCA. Notice that the common approach of taking the average over all voxel activities belonging to each RSN corresponds to k = 1. Beyond-the-average interactions are captured by k ≥ 1. Therefore, our method can be considered as a generalization for the average activity approach.

Recently, the point process analysis described in [Tagliazucchi et al., 2012] showed that the relevant information in resting-state fMRI can be obtained by looking into discrete events resulting in relatively large amplitude BOLD signal peaks. Following this idea, we have considered the preprocessed resting fMRI time-series to be spontaneous event-related and individuated point processes corresponding to signal fluctuations with a given signature, extracting a voxel-specific HRF to be used for deconvolution.

We want to remark that the use the HRF deconvolution, apart from being conceptually mandatory in our opinion, it is also crucial for our results. Indeed we repeated our analysis while omitting the deconvolution stage, and we found a clear different pattern of IF between RSNs. In particular, IF differences were only significant for k=1 and AD showed a decreased IF w.r.t. healthy subjects (figure S3). We also verified that, in general, the use of HRF deconvolution increased the IF for all subjects in comparison to no deconvolution. Thus, if we had omitted the HRF deconvolution stage, we would have not observed the relevant role of the second and the third principal components in shaping the differences of IF in AD and controls.

Previous work has addressed DFC between the different RSNs; for instance, the authors in [Demirci et al., 2006] ICA-extracted the time-courses of spatially independent components and found differences in DFC between schizophrenia and control conditions. It is worth mentioning that a method for effective connectivity inference, combining PCA and GC, was proposed in [Zhou et al., 2009]. The main differences between our method (tailored to analyze the IF between RSNs) and the one developed in [Zhou et al., 2009] are the blind deconvolution with HRF and the fact that we use the number of components as a parameter to be systematically increased (up to finding statistically significant different causalities), and both driver and target ROIs are described with the same number of components. In [Zhou et al., 2009], instead, all voxels in the activated brain regions were taken as target and the PCA analysis was applied only to the driver region and not to the target one.

Following previous work extending the use of DFC to the multivariate situation [Barnett et al., 2009, Deshpande et al., 2009, Barrett et al., 2010, Liao et al., 2011], we have applied here a multivariate DFC approach to the study of the interaction between RSNs. Specifically to AD, causal interactions among the different RSNs were addressed in [Liu et al., 2012] by a multivariate Granger causality. The authors found an increase in IF between RSNs in relation to the default mode network and the executive control one, which is in agreement with the increase of IF reported here, possibly suggesting compensatory processes in the brain networks underlying AD. Similarly and more recent, an increase of connectivity in the default mode network was found by other authors in [Liang et al., 2014] in amnesic Mild Cognitive Impairment by using GC.

The main result of the present study is the finding that the AD condition had higher IF between RSNs in comparison to control. This is apparently in contradiction to a recent paper [Li et al., 2013] where a Bayesian network approach reported a general decrease in connectivity strength for AD. Moreover, the authors in [Li et al., 2013] found an increase in connectivity between the default mode network and the dorsal attention network when using the average voxels activity, that in our approach, this situation is equivalent to considering k=1. Here, if we apply the HRF deconvolution preprocessing, we find no significant differences between controls and AD at k=1; but, if no deconvolution is applied, k=1 for AD shows a significantly decreased IF w.r.t. healthy subjects, in fully agreement with [Li et al., 2013]. Moreover, for k=1 we find an increase of the interaction from the default mode network to the dorsal visual (left) network and this is also consistent with the findings in [Li et al, 2013]. It follows that the results in the paper [Li et al., 2013] are consistent with the application of our method when just one component is considered and the HRF deconvolution is omitted. Our findings suggest that in AD the second (and, to a lesser extent, the third) components of the signals within RSNs are responsible of the increased IF. Approximating each RSN by a single signal is not enough to put in evidence these phenomena.

The IF increase found in AD might have different causes; perhaps, due to a compensatory re-organization of brain circuits due to synaptic plasticity [Adams, 1991], or due to the fact that AD patients might fail in ignoring irrelevant inputs when integrating information to performing particular cognitive tasks [Rodriguez et al., 1999], or due the reduction in inhibitory modulatory influence across the whole-brain network in AD [Amieva et al., 2004, Bentley et al., 2008, Rytsar et al., 2011], but the exact mechanism producing an increase of IF in AD needs for further investigation.

The small population size included in this study only allows for a limited interpretation of the details of connectivity changes between networks. However, a conservative approach to this analysis indicates that the IF implicated in sensory processing networks and the DMN is relatively increased in AD compared to controls. Both reductions as well as increments in functional connectivity have been previously reported between brain regions in early AD. Interestingly, most of the regional connectivity increments have been described in the frontal regions, overlapping with regions belonging to the executive control network, DMN and frontal regions of the dorsal visual processing stream [Sanz-Arigita et al., 2010].

## Acknowledgments

We acknowledge Prof. Dante R. Chialvo who first motivated us to study information flow between resting state networks and financial support from Ikerbasque: The Basque Foundation for Science, Gobierno Vasco (Saiotek SAIO13-PE13BF001), Junta de Andalucia (P09-FQM-4682) and Euskampus at UPV/EHU to JMC; Ikerbasque Visiting Professor at Biocruces and project "BRAhMS – Brain Aura Mathematical Simulation" (AYD-000-285), co-funded by Bizkaia Talent and European Commission through COFUND programme to SS; pre-doctoral contract from the Basque Government, Eusko Jaurlaritza, grant PRE_2014_1_252, to AE.

*ADNI Acknowledgments:*

Data collection and sharing for this project was funded by the Alzheimer's Disease Neuroimaging Initiative (ADNI) National Institutes of Health grant U01 AG024904. ADNI is funded by the National Institute on Aging, the National Institute of Biomedical Imaging and Bioengineering, and through generous contributions from the following: Abbott, AstraZeneca AB, Amorfix, Bayer Schering Pharma AG, Bio- clinica Inc., Biogen Idec, Bristol-Myers Squibb, Eisai Global Clinical Development, Elan Corporation, Genentech, GE Healthcare, Innogenetics, IXICO, Janssen Alzheimer Immunotherapy, Johnson and Johnson, Eli Lilly and Co., Medpace, Inc., Merck and Co., Inc., Meso Scale Diagnostic, & LLC, Novartis AG, Pfizer Inc, F. Hoffman-La Roche, Servier, Synarc, Inc., and Takeda Pharmaceuticals, as well as non-profit partners the Alzheimer's Association and Alzheimer's Drug Discovery Foundation, with participation from the U.S. Food and Drug Administration. Private sector contributions to ADNI are facilitated by the Foundation for the National Institutes of Health (www.fnih.org). The grantee organization is the Northern California Institute for Research and Education, and the study is coordinated by the Alzheimer's Disease Cooperative Study at the University of California, San Diego. ADNI data are disseminated by the Laboratory for Neuro Imaging at the University of California, Los Angeles. This research was also supported by NIH grants P30 AG010129, K01 AG030514, and the Dana Foundation.

Zhou Z, Chen Y, Ding M, Wright P, Lu Z, and Liu Y. Analyzing brain networks with PCA and conditional Granger causality. Hum Brain Mapp 30:2197–2206, 2009.

**List of captions:**

**Figure 1: Methodological sketch.** Red rectangles are remarking key stages in our approach.

**Figure 2: Average transferred information between all RSNs as a function of the number of principal components.** Control (top) vs AD (bottom). The pattern of transferred information is the same for the two conditions; it increases from k = 1 up to the maximum at k = 5 to start to decrease up to zero information for k ≥ 10. This means that the k = 5, multivariate, IF between the different RSNs is most informative than in any other dimension. * represent statistical differences between control and Alzheimer, p = 0.05 (Bonferroni correction). Standard error (depicted in red) has been calculated across subjects for each group, control (n=10) vs Alzheimer's (n=10). Information has been calculated in nats (i.e., Shannon entropies have been calculated in natural logarithms); but to transform to information bits we have to multiply the value in nats by 1.44.

**Figure 3: Networks of IF between the different RSNs.** For k = 2 (occurring the biggest difference between control and AD in figure 2), we have represented the multivariate IF between the different RSNs. Control (left) vs AD (right) for the two directions of IF (top and bottom). IF values are proportional to arrow thickness. Values represented in figure 2 are the average among all the arrows represented in this figure taking into account the two flow directions (top and bottom). Only for visualization purposes, values of IF have been normalized to the common maximum (marked with the red arrow), corresponding to TE = 0.078 nats from the executive control network to the medial visual (left) in the AD condition. Dashed arrow from the sensory-motor network to the medial visual corresponds to the minimum value, that before normalization was TE=0.006, and after normalization was fixed to zero.

**Figure 4: Control minus AD differences in the total IF per RSN.** Outward information (blue) and inward information (red) from/to each different RSN. Error bars have been calculated across subjects for each group. Notice that values in this figure are much higher than those in figure 2 due to two reasons: first, values in figure 2 corresponds to the average value of IF, that taking off principal diagonal elements, this implied to divide each IF value by a factor of 56. Second, because to calculate both outward and inward information, we sum over columns and rows respectively, this means to multiply each IF value by a factor 6 (not including the self-node information, neither the element in the principal diagonal). Thus, values in this figure might be even up to 336 times bigger.

**Figure 5: Brain maps of statistical significance localizing the k = 2 component within each RSN.** After a two-sample unpaired t-test (see methods), we are representing two possible contrasts: in red, the figure shows the significant activity existing in AD but nonexistent in control. In blue, vice versa, differences which exist in control but not in AD.

**List of Tables:**

**Table 1: AD patients**

| ADNI ID | sex | age |
|---|---|---|
| 002 S 5018 | M | 73 |
| 006 S 4867 | M | 75 |
| 018 S 4696 | F | 73 |
| 018 S 5074 | F | 75 |
| 100 S 5106 | M | 74 |
| 130 S 4641 | F | 74 |
| 130 S 4984 | F | 73 |
| 130 S 5059 | M | 73 |
| 136 S 4993 | F | 72 |
| 006 S 4546 | M | 72 |

**Table 2: healthy subjects**

| ADNI ID | sex | age |
|---|---|---|
| 006 S 4485 | M | 73 |
| 006 S 4150 | M | 75 |
| 002 S 4262 | F | 73 |
| 002 S 4270 | F | 75 |
| 012 S 4026 | M | 74 |
| 002 S 4264 | F | 74 |
| 018 S 4349 | F | 73 |
| 018 S 4400 | M | 72 |
| 031 S 4032 | F | 72 |
| 002 S 4225 | M | 72 |

**Table 3: RSN-masks size**

| RSN name | Number of Voxels |
|---|---|
| medial visual | 5649 |
| lateral visual | 8470 |
| auditory | 10894 |
| sensory-motor | 7668 |
| default mode | 8201 |
| executive control | 15209 |
| dorsal visual right | 12197 |
| dorsal visual left | 10524 |

**Table 4: Localization of brain differences in AD vs control using the AAL parcellation**

| RSN | Regions in AD but not in control (red in figure 5) | Regions in control but not in AD (blue in figure 5) |
|---|---|---|
| Medial Visual | Calcarine sulcus, Cuneus, Lingual Gyrus | Paracentral Lobule |
| Lateral Visual | Middle Occipital Gyrus | Superior Occipital Gyrus, Middle Occipital Gyrus |
| Auditory | Thalamus | Superior Temporal Gyrus |
| Sensory-Motor | Rolandic Operculum, Heschl's Gyrus, Superior Temporal Gyrus | Precentral |
| Default Mode | Medial Frontal Gyrus, Thalamus | Midcingulate Area, Cuneus, Angular Gyrus |
| Executive Control | Supramarginal Gyrus | Superior Frontal Gyrus, Middle Frontal Gyrus |
| Dorsal Visual Right | Inferior Parietal Lobule, Angular Gyrus | Inferior Frontal Gyrus Pars Triangularis, Insula |
| Dorsal Visual Left | Angular Gyrus | --none upper 15% overlapping-- |

**Figure 1:**

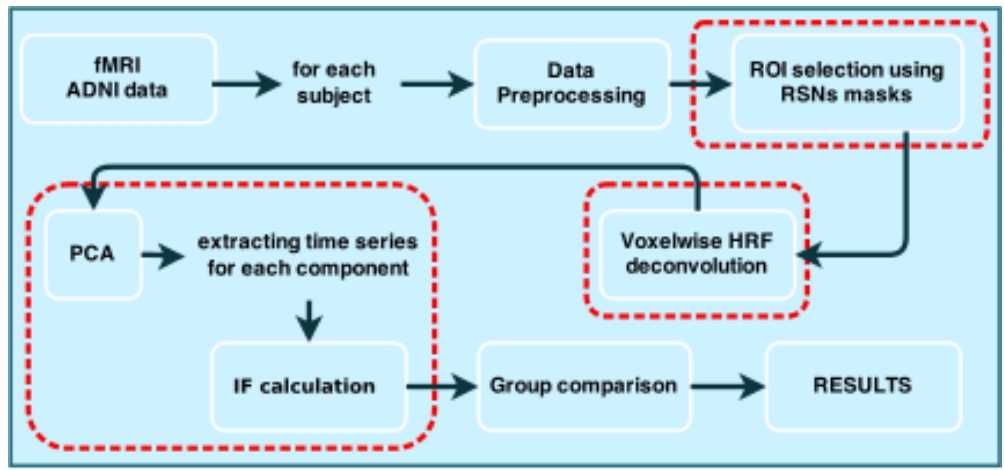

**Figure 2:**

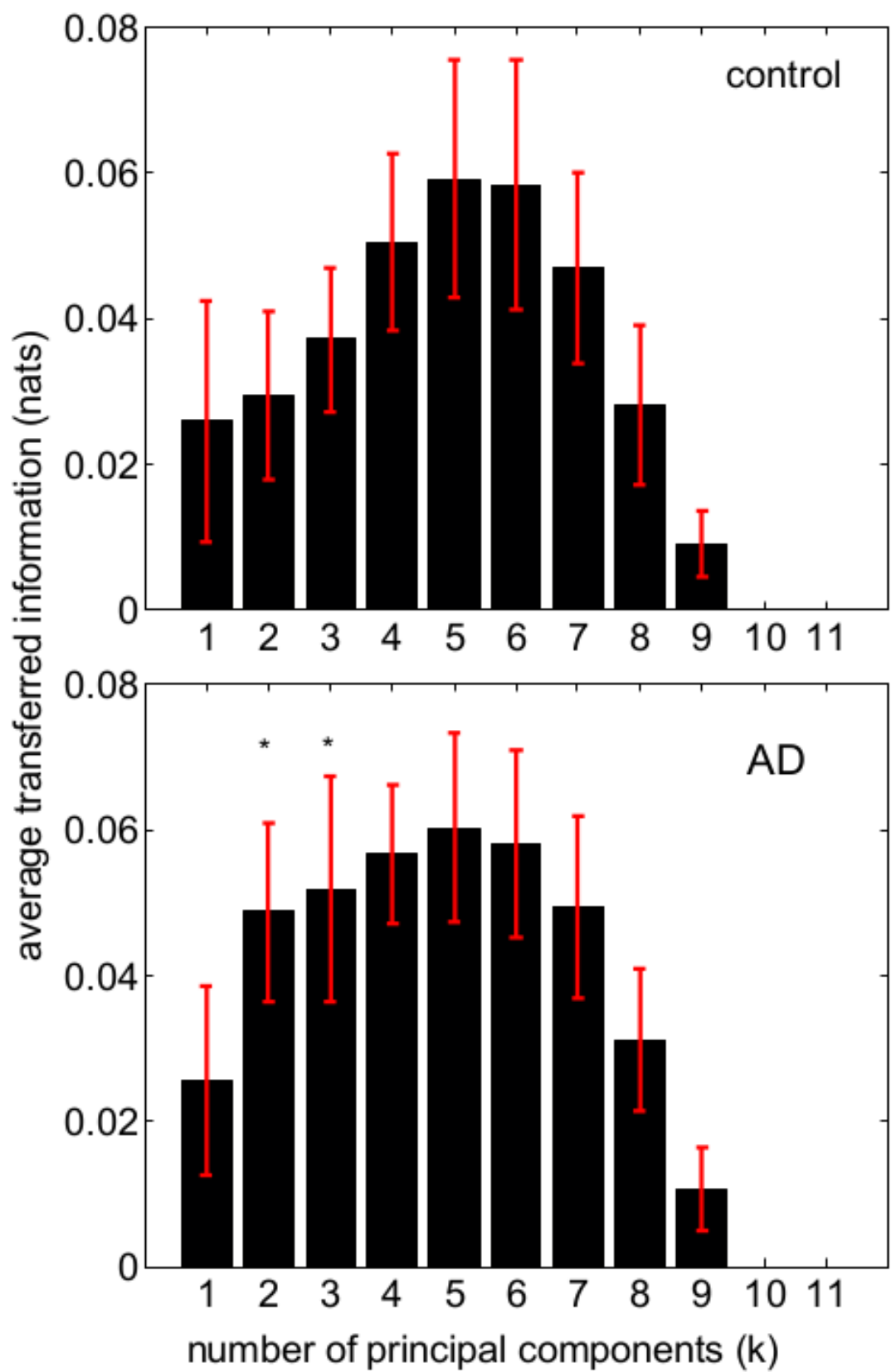

**Figure 3:**

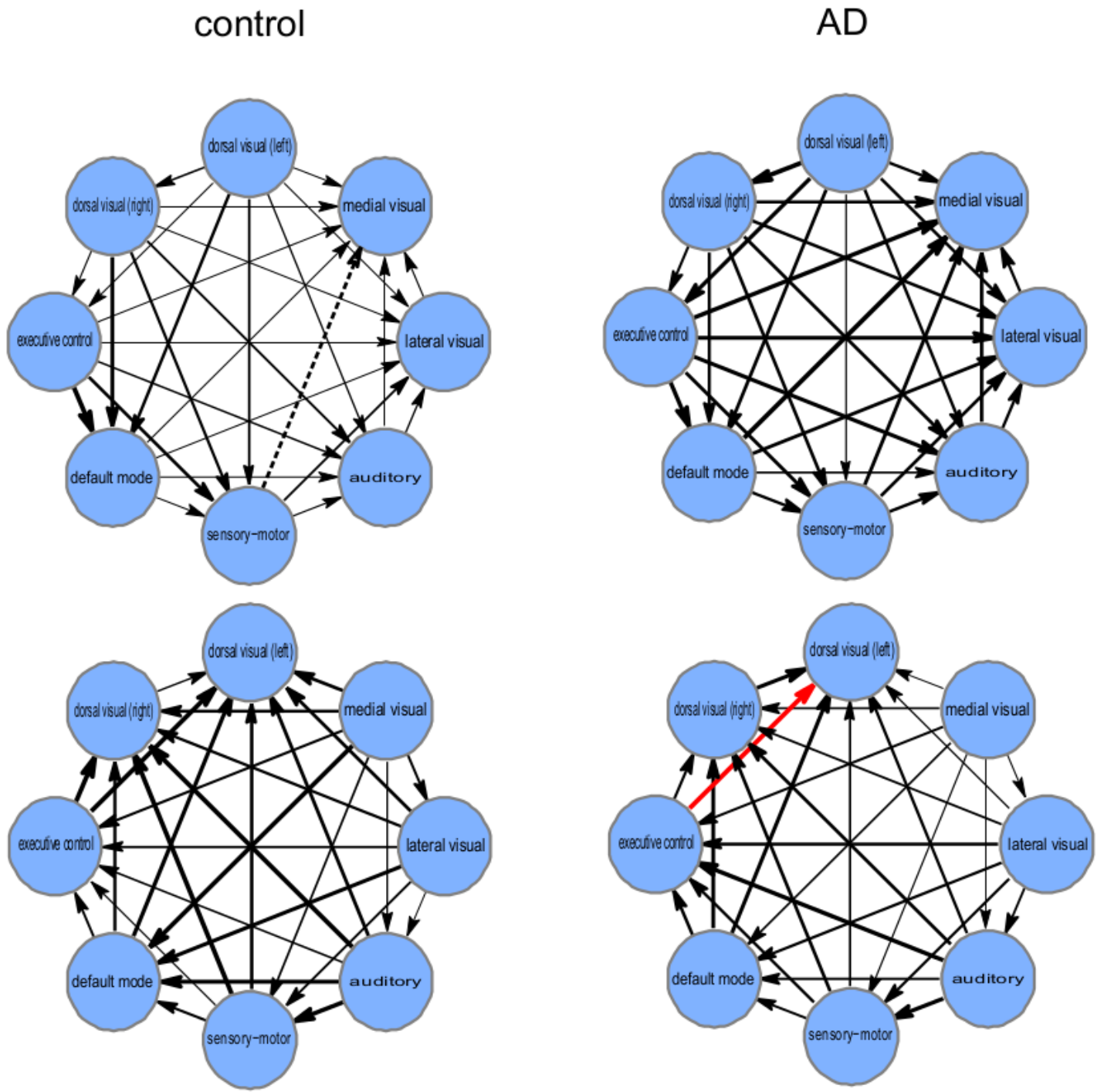

**Figure 4:**

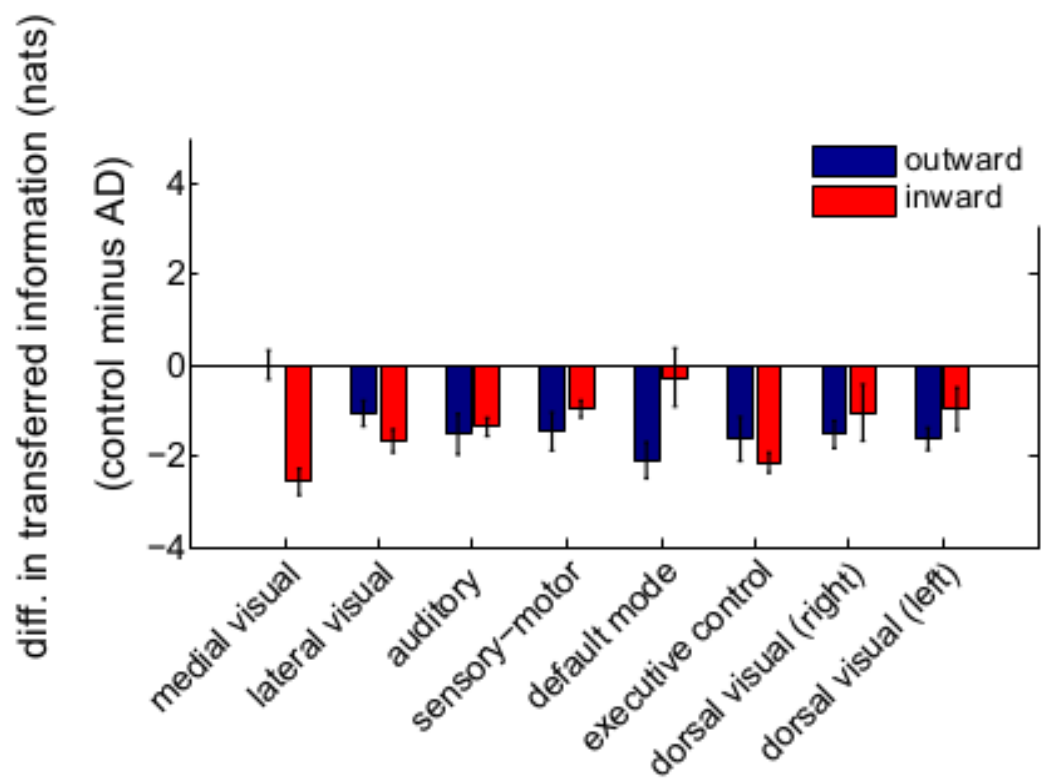

**Figure 5:**

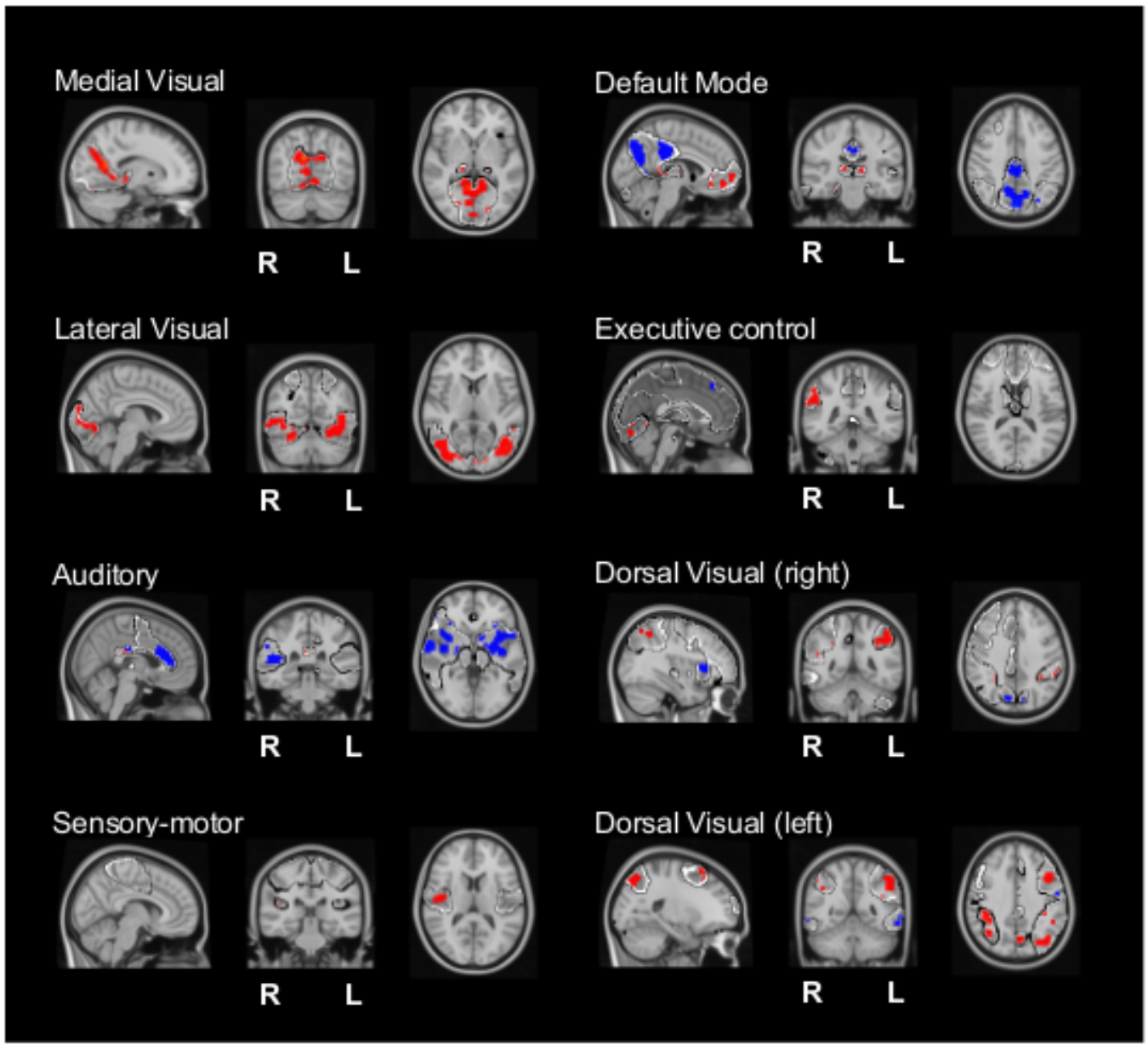

**List of captions of supplementary figures:**

**Figure S1: Amount of data variability captured for different principal components, from k=1 to k=10.**

**Figure S2: Similar to figure S1, but for k varying from 1 to 80.**

**Figure S3: Similar to figure 2, but omitting the deconvolution stage. *** represent statistical differences between control and Alzheimer, p = 0.05 (Bonferroni correction).

**Figure S1:**

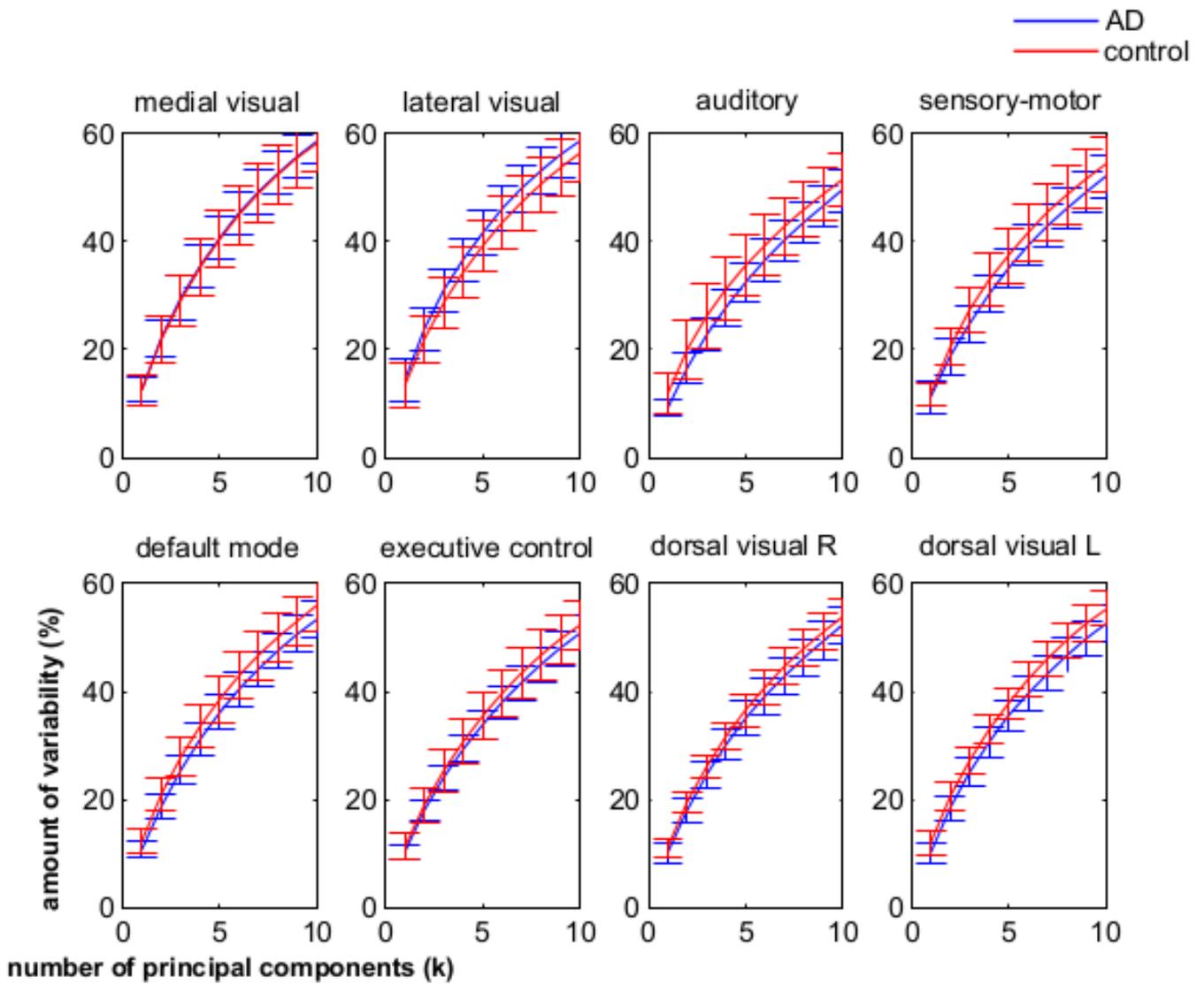

**Figure S2:**

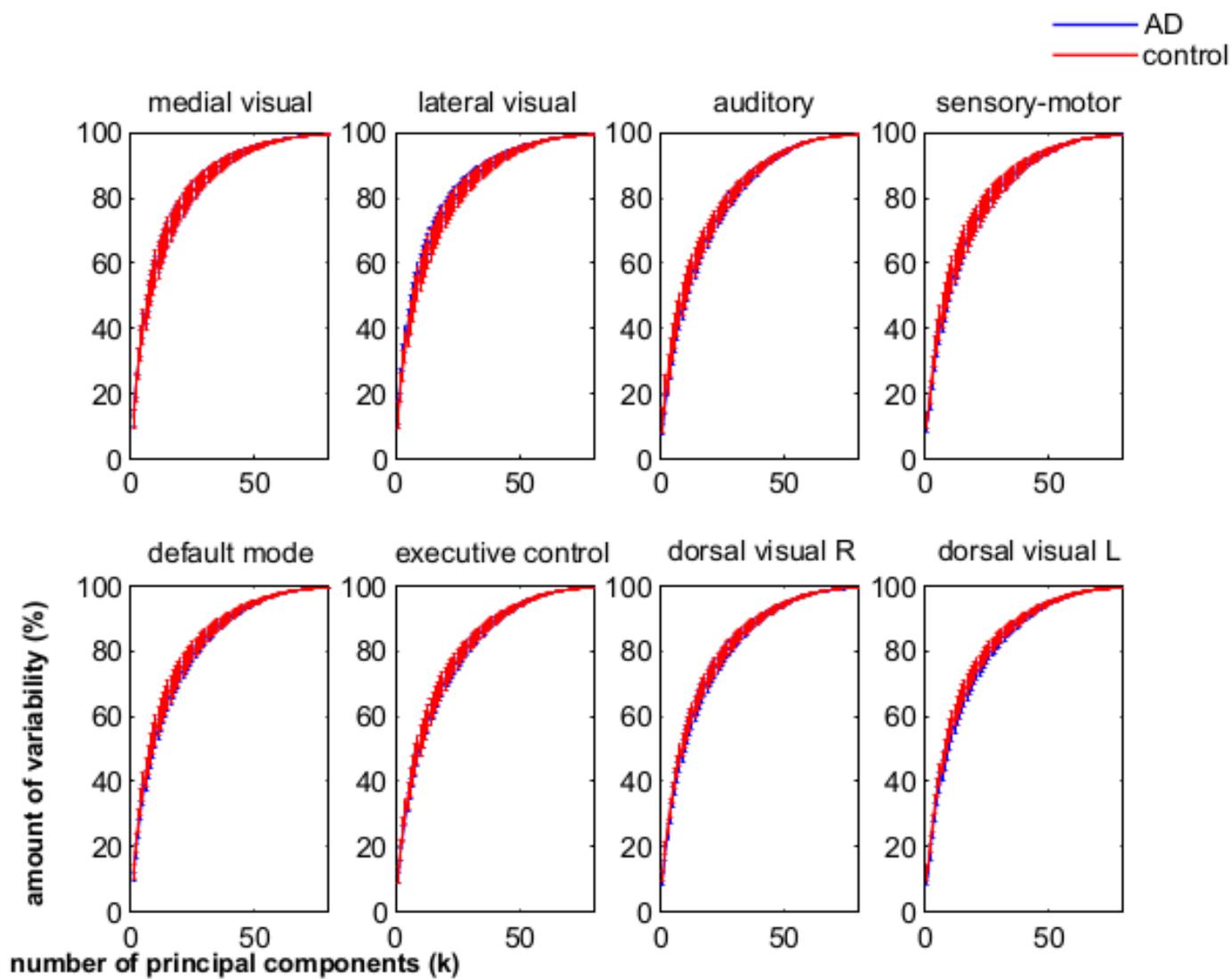

**Figure S3:**

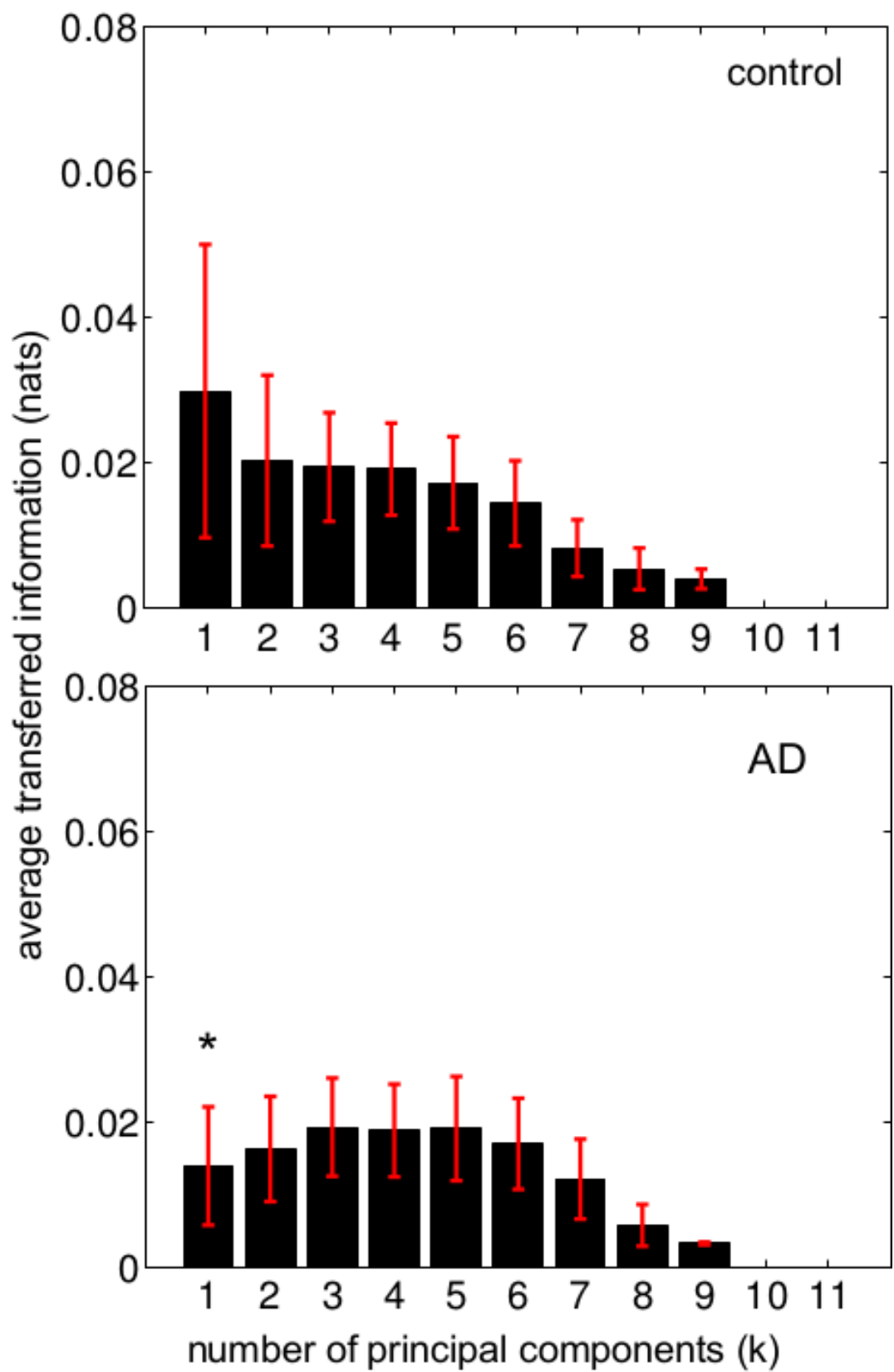